\documentclass[epj,referee]{svjour}
\usepackage{graphics}
\begin{document}
\title{Electronic structure and magnetic properties of correlated metals}
\subtitle{A local self-consistent perturbation scheme}
\author{M. I. Katsnelson\inst{1,2,3} \and A. I. Lichtenstein\inst{1}
}                     
\institute{
 University of Nijmegen, NL 6525 ED Nijmegen, The Netherlands 
\and Department of Physics, Uppsala University, Box 530, 751 21 Uppsala, Sweden 
\and Institute of Metal Physics, 620219 Ekaterinburg , Russia}
\date{Received: date / Revised version: date}
%
\abstract{
In the framework of {\it ab initio} dynamical mean field theory for
realistic electronic structure calculations
a new perturbation scheme which combine the $T$-matrix and
fluctuating exchange approximations has been proposed.
This method is less computationally expensive
than  numerically exact quantum Monte Carlo technics and
give an adequate description of the
electronic structure and exchange interactions for magnetic metals.
We discuss a simple expression for the exchange interactions
corresponding to the neglecting of the vertex corrections which
becomes exact for the spin-wave stiffness in the local approximation.
Electronic structure, correlation effects and exchange interactions
for ferromagnetic nickel have been discussed.
} 
\maketitle
\section{Introduction}

Electronic structure and magnetic properties of iron-group metals are a
subject of great interest for a very long period (for review of early
theories see \cite{herring,vons,moriya}). Density functional (DF) theory in
the form of local spin density approximation (LSDA) or generalized gradient
approximation (GGA), which is the base of modern microscopic theory of
solids, is faced with a series of difficulties when describing the
photoemission, thermoemission and other spectra of Fe and Ni as well as
their finite-temperature magnetic properties (see \cite
{IKT,vollhardt,book,spflex,PRL} and Refs. therein). The electron correlation
effects should be taken into account to solve these problems. There were a
lot of attempts to include these effects in band structure calculations of
transition metals \cite{liebsch,treglia,manghi,nolting,steiner,drchal}. Probably
the most accurate and successful approach is the use of the dynamical
mean-filed theory (DMFT, \cite{DMFT,GKKR}) with the solution of effective
impurity problem by numerically exact QMC methods; it has been applied to the
magnetism of transition metals in Refs.\cite{exchplus,PRL}. Unfortunately,
this approach is very cumbersome and expensive computationally; besides
that, the QMC method deals with the ``truncated'' two-indices interaction
matrix instead of the complete four-indices one (see \cite{exchplus}).
Alternatively, a scheme has been proposed in Ref.\cite{spflex} basing on a
multiband spin-polarized generalization of the ``fluctuating exchange''
(FLEX) approximation by Bickers and Scalapino \cite{flex}. Original
formulation of the FLEX approximation treats in the equal way both
particle-hole (PH) and particle-particle (PP) channels. However, their roles
in magnetism are completely different: the interaction of electrons with
spin fluctuations in PH channel leads to the most relevant correlation
effects \cite{moriya} whereas PP processes are important for the
renormalizations of the effective interactions in spirit of $T$-matrix
approach (``ladder approximation'') by Galitskii \cite{tmatr} and Kanamori
\cite{kanamori}. Therefore we used in Ref. \cite{spflex} a ``two-step''
procedure when, at first, bare matrix vertex is replaced by $T$-matrix, and,
secondly, PH channel processes with this effective interaction are taken
into account explicitly. Note that the first attempt to combine the $T$-matrix
and particle-hole correlations in relation with the problem of the magnetism of
transition metals has been done by Liebsch {\cite{liebsch}) but in a different way
(introducing of the particle-hole renormalization into the $T$-matrix which is opposite
in this sense to the approach of Ref. \cite{spflex}).

The latter (``two-step-FLEX'') approximation has high enough accuracy
for the Hubbard model \cite{fleck}, as well as for real systems with moderate
correlations $U < W/2$ where $U$ is the Hubbard on-site repulsion
energy and $W$ is the bandwidth \cite{spflex}. The replacement of the
bare Coulomb interaction by the $T$-matrix can be justified accurately,
at least for the spin-wave temperature region, both for the Hubbard
model \cite{IK90} and for the s-d exchange (spin-fermion) model \cite{IK85}.
However, specific form of the approximation used in \cite{spflex} can be
improved further by taking into account the spin-dependence of the $T$-matrix.
Here we present the formulation of the spin-polarized T-matrix-FLEX (SPTF)
approximation and its application to the electron structure of ferromagnetic nickel.

\section{Computational approach}

We start with the general many-body Hamiltonian in the LDA+U scheme \cite
{revU}:
\begin{eqnarray}
H &=&H_{t}+H_{U}  \nonumber \\
H_{t} &=&\sum\limits_{\lambda \lambda ^{\prime }\sigma }t_{\lambda \lambda
^{\prime }}c_{\lambda \sigma }^{+}c_{\lambda ^{\prime }\sigma }  \nonumber \\
H_{U} &=&\frac{1}{2}\sum\limits_{\left\{ \lambda _{i}\right\} \sigma \sigma
^{\prime }}\left\langle \lambda _{1}\lambda _{2}\left| v\right| \lambda
_{1}^{\prime }\lambda _{2}^{\prime }\right\rangle c_{\lambda _{1}\sigma
}^{+}c_{\lambda _{2}\sigma ^{\prime }}^{+}c_{\lambda _{2}^{\prime }\sigma
^{\prime }}c_{\lambda _{1}^{\prime }\sigma \,,}  \label{hamilU}
\end{eqnarray}
where $\lambda =im$ are the site number $\left( i\right) $ and orbital $%
\left( m\right) $ quantum numbers, $\sigma =\uparrow ,\downarrow $ is the
spin projection, $c^{+},c$ are the Fermion creation and annihilation
operators, $H_{t}$ is the effective single-particle Hamiltonian from the
LDA, corrected for the double-counting of average interactions among
correlated electrons as it will be described below, and the Coulomb matrix
elements are defined in the standard way
\begin{equation}
\left\langle 12\left| v\right| 34\right\rangle =\int d{\bf r}d{\bf r}%
^{\prime }\psi _{1}^{\ast }({\bf r})\psi _{2}^{\ast }({\bf r}^{\prime
})v\left( {\bf r-r}^{\prime }\right) \psi _{3}({\bf r})\psi _{4}({\bf r}%
^{\prime }),  \label{coulomb}
\end{equation}
where we define for briefness $\lambda _{1}\equiv 1$ etc. Following Ref. \cite
{tmatr} we take into account the ladder ($T$-matrix) renormalization of
the effective interaction:
\begin{eqnarray}
&&\left\langle 13\left| T^{\sigma \sigma ^{\prime }}\left( i\Omega \right)
\right| 24\right\rangle = \left\langle 13\left| v\right| 24\right\rangle -%
\frac{1}{\beta }\sum\limits_{\omega }\sum\limits_{5678}\left\langle 13\left|
v\right| 57\right\rangle * \nonumber \\ 
&&G_{56}^{\sigma }\left( i\omega \right)
G_{78}^{\sigma ^{\prime }}\left( i\Omega -i\omega \right) \left\langle
68\left| T^{\sigma \sigma ^{\prime }}\left( i\Omega \right) \right| 24\right\rangle ,
\label{tmatrix}
\end{eqnarray}
where $\omega =(2n+1)\pi T$ are the Matsubara frequencies for temperature $%
T\equiv \beta ^{-1}$ ($n=0,\pm 1,...$). Further we rewrite the perturbation
theory in terms of this effective interaction matrix.

At first, we take into account the ``Hartree'' and ``Fock'' diagrams with
the replacement of the bare interaction by the $T$-matrix
\begin{eqnarray}
\Sigma _{12,\sigma }^{\left( TH\right) }\left( i\omega \right) &=&\frac{1}{%
\beta }\sum\limits_{\Omega }\sum\limits_{34\sigma ^{\prime }}\left\langle
13\left| T^{\sigma \sigma ^{\prime }}\left( i\Omega \right) \right|
24\right\rangle G_{43}^{\sigma ^{\prime }}\left( i\Omega -i\omega \right)
\nonumber \\
\Sigma _{12,\sigma }^{\left( TF\right) }\left( i\omega \right) &=&-\frac{1}{%
\beta }\sum\limits_{\Omega }\sum\limits_{34}\left\langle 14\left| T^{\sigma
\sigma }\left( i\Omega \right) \right| 32\right\rangle G_{34}^{\sigma
}\left( i\Omega -i\omega \right) \nonumber \\ \label{HarFock}
\end{eqnarray}
Note that $\Sigma ^{(TH)}+$ $\Sigma ^{(TF)}$ contains exactly all the
second-order contributions as it can be easily seen from the corresponding
Feynman diagrams. Now we have to consider the contribution of particle-hole
excitations to the self-energy. Similar to \cite{spflex} we will replace in
the corresponding diagrams the bare interaction by the static limit of the $T$%
-matrix (as it was already mentioned, it can be justified by the explicit
calculation of the electron and magnon Green functions of a ferromagnet, at least,
for spin-wave temperature region \cite{IK85,IK90}). We improve the approximation
\cite{spflex} by taking into account the $T$-matrix spin dependence. When considering
the particle-hole channel we replace in the Hamiltonian (\ref{hamilU})
$v\rightarrow T^{\sigma \sigma^{\prime }}$
which is the solution of Eq.(\ref{tmatrix}) at $\Omega =0.$ Eq.
(\ref{HarFock}) is exact in the limit of low electron (or hole) density
which is important for the criterion of magnetism, e.g., in the case of nickel
(with almost completely filled $d$- band).

Now we rewrite the effective Hamiltonian (\ref{hamilU}) with the replacement
$\left\langle 12\left| v\right| 34\right\rangle $ by $\left\langle 12\left|
T^{\sigma \sigma ^{\prime }}\right| 34\right\rangle $ in $H_{U}$. To
consider the correlation effects due to PH channel we have to separate
density ($d$) and magnetic ($m$) channels as in Ref.\cite{flex}

\begin{eqnarray}
d_{12}&=&\frac 1{\sqrt{2}}\left( c_{1\uparrow }^{+}c_{2\uparrow
}+c_{1\downarrow }^{+}c_{2\downarrow }\right) \nonumber \\
m_{12}^{0} &=&\frac{1}{\sqrt{2}}\left( c_{1\uparrow }^{+}c_{2\uparrow
}-c_{1\downarrow }^{+}c_{2\downarrow }\right)  \nonumber \\
m_{12}^{+} &=&c_{1\uparrow }^{+}c_{2\downarrow }  \nonumber \\
m_{12}^{-} &=&c_{1\downarrow }^{+}c_{2\uparrow }\,,  \label{chan}
\end{eqnarray}
Then the interaction Hamiltonian can be rewritten in the following matrix
form
\begin{equation}
H_{U}=\frac{1}{2}Tr\left( D^{+}\ast V^{\parallel }\ast D+m^{+}\ast
V_{m}^{\perp }\ast m^{-}+m^{-}\ast V_{m}^{\perp }\ast m^{+}\right)
\label{hamnew}
\end{equation}
where $\ast $ means the matrix multiplication with respect to the pairs of
orbital indices, e.g.
\begin{eqnarray*}
\left( V_{m}^{\perp }\ast m^{+}\right) _{11^{\prime }}
&=&\sum\limits_{34}\left( V_{m}^{\perp }\right) _{11^{\prime },22^{\prime
}}m_{22^{\prime }}^{+},
\end{eqnarray*}
the supervector D defined as
\begin{eqnarray*}
D &=&\left( d,m^{0}\right) ,  D^{+}=\left(
\begin{array}{c}
d^{+} \\
m_{0}^{+}
\end{array}
\right) ,
\end{eqnarray*}
and the effective interactions have the following form:
\begin{eqnarray}
&&\left( V_{m}^{\perp }\right) _{11^{\prime },22^{\prime }} =-\left\langle
12\left| T^{\uparrow \downarrow }\right| 2^{\prime }1^{\prime }\right\rangle
 \nonumber \\
&&V^{\parallel } =\left(
\begin{array}{cc}
V^{dd} & V^{dm} \\
V^{md} & V^{dd}
\end{array}
\right)   \nonumber \\
&&V_{11^{\prime },22^{\prime }}^{dd} =\frac{1}{2}\sum\limits_{\sigma \sigma
^{\prime }}\left\langle 12\left| T^{\sigma \sigma ^{\prime }}\right|
1^{\prime }2^{\prime }\right\rangle -\frac{1}{2}\sum\limits_{\sigma
}\left\langle 12\left| T^{\sigma \sigma }\right| 2^{\prime }1^{\prime
}\right\rangle   \nonumber \\
&&V_{11^{\prime },22^{\prime }}^{mm} =\frac{1}{2}\sum\limits_{\sigma \sigma
^{\prime }}\sigma \sigma ^{\prime }\left\langle 12\left| T^{\sigma \sigma
^{\prime }}\right| 1^{\prime }2^{\prime }\right\rangle -\frac{1}{2}%
\sum\limits_{\sigma }\left\langle 12\left| T^{\sigma \sigma }\right|
2^{\prime }1^{\prime }\right\rangle   \nonumber \\
&&V_{11^{\prime },22^{\prime }}^{dm}=V_{22^{\prime },11^{\prime }}^{md}= \nonumber \\
&&\frac{1}{2}[\left\langle 12\left| T^{\uparrow \uparrow }\right| 1^{\prime
}2^{\prime }\right\rangle -\left\langle 12\left| T^{\downarrow \downarrow
}\right| 1^{\prime }2^{\prime }\right\rangle - 
\left\langle 12\left|
T^{\uparrow \downarrow }\right| 1^{\prime }2^{\prime }\right\rangle + \nonumber \\
&& \left\langle 12\left| T^{\downarrow \uparrow }\right| 1^{\prime }2^{\prime
}\right\rangle 
-\left\langle 12\left| T^{\uparrow \uparrow }\right| 2^{\prime }1^{\prime
}\right\rangle + 
\left\langle 12\left| T^{\downarrow \downarrow }\right|
2^{\prime }1^{\prime }\right\rangle ]
\label{effpotent}
\end{eqnarray}
To calculate the particle-hole (P-H) contribution to the electron self-energy we first have
to write the expressions for the generalized susceptibilities, both
transverse $\chi ^{\perp }$ and longitudinal $\chi ^{\parallel }$.
The corresponding expressions are the same as in Ref.\cite{spflex} but with another
definition of the interaction vertices. One has
\begin{equation}
\chi ^{+-}(i\omega )=\left[ 1+V_m^{\perp }*\Gamma ^{\uparrow \downarrow
}(i\omega )\right] ^{-1}*\Gamma ^{\uparrow \downarrow }(i\omega )\,,
\label{xi+-}
\end{equation}
where
\begin{equation}
\Gamma _{12,34}^{\sigma \sigma ^{\prime }}\left( \tau \right)
=-G_{23}^\sigma \left( \tau \right) G_{41}^{\sigma ^{\prime }}\left( -\tau
\right)  \label{gamma}
\end{equation}
is an ``empty loop'' susceptibility and $\Gamma (i\omega )$ is its Fourier transform,
$\tau $ is the imaginary time. The
corresponding longitudinal susceptibility matrix has a more complicated form:
\begin{equation}
\chi ^{\parallel }(i\omega )=\left[ 1+V^{\parallel }*\chi _0^{\parallel
}(i\omega )\right] ^{-1}*\chi _0^{\parallel }(i\omega ),  \label{xipar}
\end{equation}
and the matrix of bare longitudinal susceptibility is
\begin{equation}
\chi _{0}^{\parallel }=\frac{1}{2}\left(
\begin{array}{cc}
\Gamma ^{\uparrow \uparrow }+\Gamma ^{\downarrow \downarrow }\, & \,\Gamma
^{\uparrow \uparrow }-\Gamma ^{\downarrow \downarrow } \\
\Gamma ^{\uparrow \uparrow }-\Gamma ^{\downarrow \downarrow }\, & \,\Gamma
^{\uparrow \uparrow }+\Gamma ^{\downarrow \downarrow }
\end{array}
\right) ,  \label{xi0par}
\end{equation}
in the $dd$-, $dm^{0}$-, $m^{0}d$-, and $m^{0}m^{0}$- channels ($d,m^{0}=1,2$
in the supermatrix indices). An important feature of these equations is the
coupling of longitudinal magnetic fluctuations and of density fluctuations.
It is not present in the one-band Hubbard model due to the absence of the interaction
of electrons with parallel spins. For this case Eqs. (\ref{xi+-},\ref{xipar})
coincides with the well-known result \cite{kubo}.

Now we can write the particle-hole contribution to the self-energy. Similar
to Ref.\cite{spflex} one has
\begin{equation}
\Sigma _{12,\sigma }^{(ph)}\left( \tau \right) =\sum\limits_{34,\sigma
^{\prime }}W_{13,42}^{\sigma \sigma ^{\prime }}\left( \tau \right)
G_{34}^{\sigma ^{\prime }}\left( \tau \right) ,  \label{sigph}
\end{equation}
with the P-H fluctuation potential matrix:

\begin{equation}
W^{\sigma \sigma ^{\prime }}\left( i\omega \right) =\left[
\begin{array}{cc}
W^{\uparrow \uparrow }\left( i\omega \right) & W^{\perp }\left( i\omega
\right) \\
W^{\perp }\left( i\omega \right) & W^{\downarrow \downarrow }\left( i\omega
\right)
\end{array}
\right] ,  \label{wpp}
\end{equation}
where the spin-dependent effective potentials are defined as
\begin{eqnarray}
W^{\uparrow \uparrow }&=&\frac{1}{2}V^{\parallel }\ast \left[ \chi ^{\parallel
}-\chi _{0}^{\parallel }\right] \ast V^{\parallel }  \nonumber \\ 
W^{\downarrow \downarrow }&=&\frac{1}{2}V^{\parallel }\ast \left[ \widetilde{%
\chi }^{\parallel }-\widetilde{\chi }_{0}^{\parallel }\right] \ast
V^{\parallel }  \nonumber \\
W^{\uparrow \downarrow }&=&V_{m}^{\perp }\ast \left[ \chi ^{+-}-\chi _{0}^{+-}%
\right] \ast V_{m}^{\perp }  \nonumber \\
W^{\downarrow \uparrow }&=&V_{m}^{\perp }\ast \left[ \chi ^{-+}-\chi _{0}^{-+}%
\right] \ast V_{m}^{\perp }.
\end{eqnarray}
where $\widetilde{\chi }^{\parallel },\widetilde{\chi }_{0}^{\parallel }$
differ from $\chi ^{\parallel },\chi _{0}^{\parallel }$ by the replacement
of $\Gamma ^{\uparrow \uparrow }\Leftrightarrow \Gamma ^{\downarrow
\downarrow }$ in Eq.(\ref{xi0par}). We have subtracted the second-order
contributions since they have already been taken into account in Eq.(\ref{HarFock}%
).

Our final expression for the self energy is
\begin{equation}
\Sigma =\Sigma ^{(TH)}+\Sigma ^{(TF)}+\Sigma ^{(PH)}  \label{sigmatot}
\end{equation}
This formulation takes into account accurately spin-polaron effects because
of the interaction with magnetic fluctuations \cite{IK90,UFN}, the energy dependence
of $T$-matrix which is important for describing the satellite effects in Ni
\cite{liebsch}, contains exact second-order terms in $v$ and is rigorous
(because of the first term) for almost filled or almost empty bands.

The FLEX approximation can be used in principle directly to the crystal
problem taking into account the momentum dependence of the self-energy,
which would lead to very cumbersome calculations. 
To overcome this computational problem, we use
as in Ref.\cite{spflex} a local
approximation to the self-energy, corresponding to 
combination of the SPTF approach presented above with the
DMFT theory. The latter maps the many-body system onto a multi-orbital
quantum impurity, i.e. a set of local degrees of freedom in a bath described
by the Weiss field function ${\cal G}$. The impurity action (here ${\bf c}%
(\tau )=[c_{m\sigma }(\tau )]$ is a vector of Grassman variables) is given
by:
\begin{eqnarray}
S_{eff}&=&\int_{0}^{\beta }d\tau \int_{0}^{\beta }d\tau ^{\prime }Tr[{\bf c}%
^{+}(\tau ){{\bf {\cal G}}}^{-1}(\tau ,\tau ^{\prime }){\bf c}(\tau ^{\prime
})]+  \nonumber \\
&&\int_{0}^{\beta }d\tau H_{U}\left[ {\bf c}^{+}(\tau ),{\bf c}(\tau )%
\right]  \label{path}
\end{eqnarray}
It describes the spin, orbital, energy and temperature dependent
interactions of particular magnetic $3d$-atom with the rest of the crystal
and is used to compute the local Greens function matrix:
\begin{equation}
{\bf G}_{\sigma }(\tau -\tau ^{\prime })=-\frac{1}{Z}\int D[{\bf c},{\bf c}%
^{+}]e^{-S_{eff}}{\bf c}(\tau ){\bf c}^{+}(\tau ^{\prime })  \label{pathint}
\end{equation}
($Z$ is the partition function) and the impurity self energy ${{\bf {\cal G}}%
}_{\sigma }^{-1}(i\omega )-{\bf G}_{\sigma }^{-1}(i\omega )={\bf \Sigma }%
_{\sigma }(i\omega )$ .

The Weiss field function ${\cal G}$ is required to obey the self consistency condition,
which restores translational invariance to the impurity model description:
\begin{equation}
{\bf G}_{\sigma }(i\omega )=\sum_{{\bf k}}[(i\omega +\mu ){\bf 1}-H({\bf k})-%
{\bf \Sigma }_{\sigma }^{dc}(i\omega )]^{-1}  \label{BZI}
\end{equation}
where $\mu $ is the chemical potential, $H({\bf k})$ is the LDA  Hamiltonian
in an orthogonal basis. The local matrix ${\bf \Sigma }_{\sigma }^{dc}$ is the
sum of two terms, the impurity self energy and a so-called ``double counting
'' correction, $E_{dc}$ which is meant to subtract the average
electron-electron interactions already included in the LDA Hamiltonian. For
metallic systems we propose the general form of dc-correction: ${\bf \Sigma }%
_{\sigma }^{dc}\left( i\omega \right) ={\bf \Sigma }_{\sigma }\left( i\omega
\right) -\frac{1}{2}Tr_{\sigma }{\bf \Sigma }_{\sigma }\left( 0\right) $ for
non-magnetic LDA Hamiltonian \cite{PRL} and ${\bf \Sigma }_{\sigma }^{dc}\left( i\omega
\right) ={\bf \Sigma }_{\sigma }\left( i\omega \right) -{\bf \Sigma }%
_{\sigma }\left( 0\right) $ for the magnetic LSDA Hamiltonian. This is motivated
by the fact that the static part of the correlation effects are already well
described in the density functional theory. Only the $d$-part of the
self-energy is presented in our calculations, therefore ${\bf \Sigma }%
_{\sigma }^{dc}=0$ for $s$- and $p$- states as well as for non-diagonal $d-s$%
,$p$ contributions.

In spirit of the DMFT approach we have to use the Weiss function ${\cal G}%
_{\sigma }$ instead of $G_{\sigma }$ in all the expressions when calculating
the self-energy on a given site. Similar to the one-band DMFT-perturbation scheme
\cite{logan,fleck} we keep the static mean-field term in the bath Green functions:
${{\bf {\cal G}}}_{\sigma }^{-1}(i\omega )={\bf G}_{\sigma }^{-1}(i\omega )+%
{\bf \Sigma }_{\sigma }(i\omega ) - {\bf \Sigma }_{\sigma }\left( 0\right)$.

\section{Electronic structure of nickel}

We have started from the non spin-polarized LDA or spin-polarized LSDA band
structure of nickel within the TB-LMTO method \cite{OKA} in the minimal $%
s,p,d$ basis set and used numerical orthogonalization to find the $H({\bf k})
$ Hamiltonian in Eq.(\ref{BZI}). We take into account of the Coulomb interactions
only between $d$-states. The correct parameterization of the $H_{U}$%
part is indeed a serious problem. For example, first-principle estimations
of average Coulomb interactions ($U$) \cite{U,steiner} lead to unreasonably
large value of order of 5-6 eV in comparison with experimental values of the
$U$-parameter in the range of 1-2 eV for iron\cite{steiner}. Semiempirical
analysis of the appropriate interaction value \cite{oles} gives $U\simeq 3$
eV. It is shown in Refs. \cite{spflex,PRL} that an adequate description of a
broad circle of the properties of Fe and Ni in the LDA+DMFT scheme is
possible when choosing $U\simeq 2 - 3$ eV.
The difficulties with an {\it ab initio} determination of
the correct value of $U$ are connected
with complicated screening problems, definitions of orthogonal orbitals in
the crystal, and contributions of the intersite interactions. In the
quasiatomic (spherical) approximation the full $U$-matrix for the $d-$shell
is determined by the three parameters $U,J$ and $\delta J$ or equivalently
by effective Slater integrals F$^{0}$, F$^{2}$ and F$^{4}$ \cite{revU}%
. For example, U= F$^{0}$, J=(F$^{2}$+F$^{4}$)/14 and we use the simplest
way of estimating $\delta J$ or F$^{4}$ keeping the ratio F$^{2}$/F$^{4}$
equal to its atomic value 0.625 \cite{revU}.

Note that the value of intra-atomic (Hund) exchange interaction $J$ is not
sensitive to the screening and approximately equals 1 eV in different
estimations\cite{U}; further we have chosen $J=1$ eV.
For the most important parameter
$U$, which defines the bare vertex matrix (Eq. (\ref{coulomb})), we took
the values $U=2$ and $3$ eV to check the dependence of the density of
states (DOS) on $U$. To find DOS we applied a Pade approximant method
\cite{pade} for the analytical continuation of the Green function
from the Matsubara frequencies to the real energy axis. To find
the self-consistent solution of the SPTF equations we used 1024 Matsubara
frequencies and the FFT-scheme with the energy cut-off at 25 eV and temperature
around 200\ K. The sum over irreducible Brillouin zone have been made with 256
${\bf k}$-points.

Comparison of the LDA density of state and the SPTF calculation with DMFT
self-consistency for the local self-energy matrix (Fig.1) shows that the latter
does reproduce the three most important characteristic features of correlation
effects for nickel: 6 eV satellite, 30\% narrowing of the $d$-bandwidth and
50\% reduction of exchange splittings in comparison with the LSDA band
structure \cite{iwan,eber,kamp,himpsel}. For $U$=2 eV the position of satellite
is reproduced quite well, while for $U$=3 eV it is shifted to the lower energies.
Note that the LDA+DMFT consideration with the QMC solution of the
effective impurity problem gives an adequate description of the electronic
structure of Ni for the choice $U=3$ eV \cite{PRL}.
The narrowing of the $d$-bandwidth in our calculations is reasonable for the both
$U$-values. The non-magnetic LDA starting Hamiltonian is better
than the LSDA one for correct description of the\ 50\% reduction of the
spin-splittings in nickel, while for magnetic LSDA Hamiltonian the the spin-splitting
in the quasiparticle DOS remains approximately the same like in the LSDA results
(Fig.1). The local magnetic moment on nickel atom is not very sensitive to the U-values
and is equal to 0.56 $\mu_B$ for U=2 eV LDA+SPTF and 0.58 $\mu_B$ for U=3 eV LSDA+SPTF calculations.

\begin{figure}
\resizebox{0.50\textwidth}{!}{%
  \includegraphics{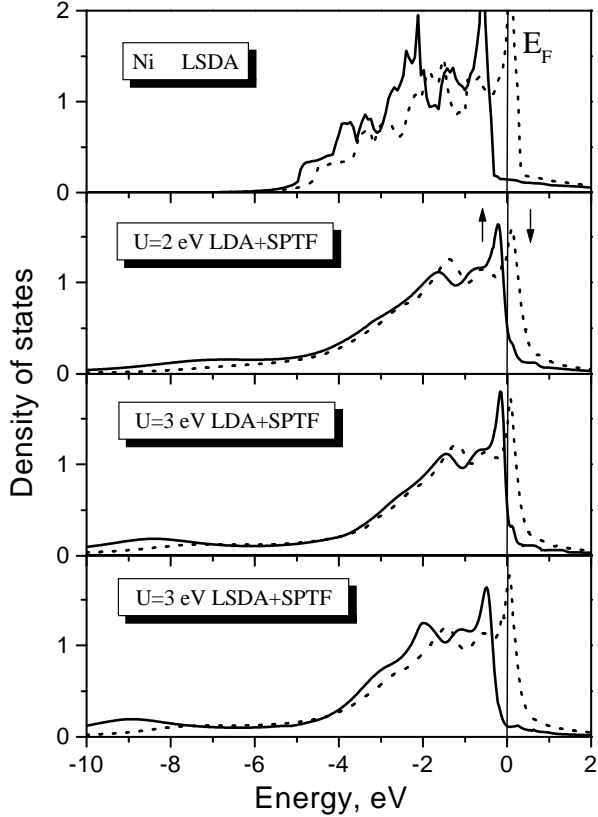}
}
\caption{
Spin-up (full lines) and spin-down (dashed lines) density of d-states
for ferromagnetic nickel in the LSDA
and the LDA+SPTF (LSDA+SPTF) calculations
for different average Coulomb interaction $U$ with $J=1$ eV
and temperature T=200~K.
}
\label{DOS}
\end{figure}
Another important correlation effect is an essential reduction of the
spin polarization near the Fermi level in comparison with the LSDA
calculations. This is connected with the spin-polaron effects because
of the mixing of the spin-up and spin-down states \cite{UFN}. They
are taken into account in our scheme due to presence of the
off-diagonal terms in the effective potential Eq.(\ref{wpp}).

\section{Exchange interactions in nickel}

Calculating the variation of the thermodynamic potential with respect to
small spin rotations with the use of the ``local force theorem'' an
effective exchange interaction parameters can be found in the following form
\cite{exchplus}
\begin{equation}
J_{ij}=-Tr_{\omega L}\left( \Sigma _{i}^{s}G_{ij}^{\uparrow }\Sigma
_{j}^{s}G_{ji}^{\downarrow }\right)   \label{Jij}
\end{equation}
where $\Sigma _{i}^{s}=\frac{1}{2}\left( \Sigma _{i}^{\uparrow }-\Sigma
_{i}^{\downarrow }\right) .$ Correspondingly, the magnon dispersion relation
$\omega _{{\bf q}}$ for a ferromagnet is defined by the formula
\begin{equation}
\omega _{{\bf q}}=\frac{4}{M}\left[ J(0)-J({\bf q})\right]   \label{om}
\end{equation}
where $M$ is the magnetic moment per unit cell, $J({\bf q})$ is the Fourier
transform of the exchange integrals defined by Eq.(\ref{Jij}). The
expression for the stiffness tensor $D_{\alpha \beta },$
\begin{equation}
\omega _{{\bf q}}=D_{\alpha \beta }q_{\alpha }q_{\beta },
\begin{array}{cc}
& {\bf q}\rightarrow 0,
\end{array}
\label{D11}
\end{equation}
reads
\begin{equation}
D_{\alpha \beta }=-\frac{2}{M}Tr_{\omega L}\sum\limits_{{\bf k}}\left(
\Sigma ^{s}\frac{\partial G^{\uparrow }\left( {\bf k}\right) }{\partial
k_{\alpha }}\Sigma ^{s}\frac{\partial G^{\downarrow }\left( {\bf k}\right) }{%
\partial k_{\beta }}\right)   \label{D21}
\end{equation}
These results generalize the LSDA expressions of Ref. \cite{LKG} to the case
of correlated systems. One can show (see Appendix) that they can be derived
using a standard diagram approach under two assumptions: (i) the locality of
the self-energy $\Sigma$ (which is fulfilled in the DMFT) and (ii) the
neglecting of the vertex corrections. The expression (\ref{D21}) for the
stiffness constant turns out to be exact in the framework of the DMFT.

We have calculated the magnon spectrum for the optimal choice $U$=2 eV and $J$=1 eV
using SPTF calculations with the non-magnetic LDA as a starting point.
The computational results are shown in Fig.2; the calculated spin-wave stiffness
constant for Ni is found to be $D=450$ meV/A$^{2}$ in an excellent agreement
with the experimental value of $455$ meV/A$^{2}$ \cite{experD}.
Note that simple approximation for exchange interactions is not allowed us to
investigate the problem of the optical mode in magnon spectrum of nickel \cite{savrasov}.
\begin{figure}
\resizebox{0.45\textwidth}{!}{%
  \includegraphics{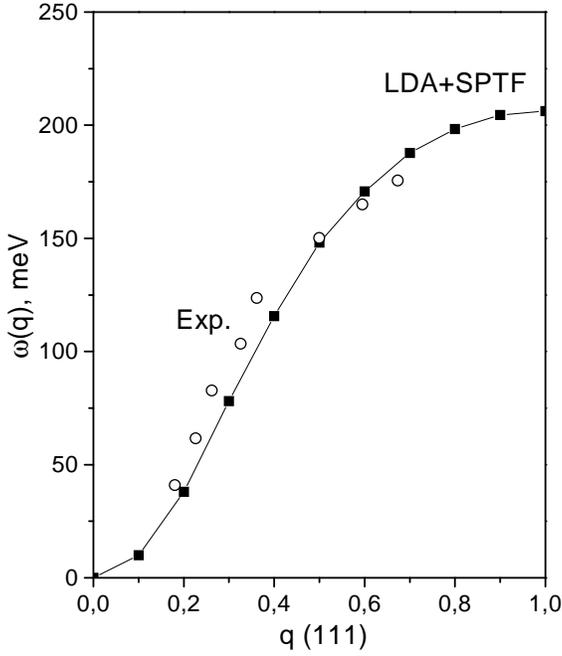}
}
\caption{
Spin-wave spectrum for ferromagnetic nickel in LDA+SPTF scheme
with $U=2$ eV and $J=1$ eV in comparison with experimental magnon spectrum (Ref. 36) in
$\Gamma -L$ direction.
}
\label{SW}
\end{figure}
\section{Conclusions}

Here we have presented the results of new SPTF approximation in the framework of
first-principle dynamical mean field theory for magnetic transition metals.
This approximation combining the $T$-matrix  and FLEX schemes gives a
satisfactory description of both electronic and magnon spectra of Ni.
In contrast with the QMC method for the solution of the effective
impurity problem, this approach, being less rigorous, is not so time-
and resource-consuming and allows to work with the most general
rotationally invariant form of the Coulomb on-site interaction.

\begin{acknowledgement}
The work was supported by the Netherlands Organization for Scientific
Research (NWO project 047-008-16) and partially supported by Russian Science Support
Foundation.
\end{acknowledgement}

\section*{Appendix: Exchange interactions and vertex corrections}

In order to elucidate the approximation behind the expression for the
exchange parameters (Eq. \ref{Jij}), we  consider the energy of a spiral
magnetic configuration with the rigid rotation of the spinor-electron
operators on site $i$ by the polar angles $\theta_{i}$ and $\varphi_{i}$:
\[c_{im}\rightarrow U\left( {\theta _{i},\varphi _{i}}\right) c_{im}\]
where
\begin{eqnarray}
U\left( {\theta ,\varphi }\right) =\left(
\begin{array}{cc}
{\cos \theta /2} & {\sin \theta /2\exp \left( {-i\varphi }\right) } \\
{-\sin \theta /2\exp \left( {i\varphi }\right) } & {\cos \theta /2}
\end{array}
\right),   \nonumber 
\end{eqnarray}
assuming that $\theta_{i} = const$ and $\varphi_{i} = {\bf qR_{i}}$ where
${\bf R_{i}}$ is the site lattice vector. Since we take into account only on-site
correlation effects the interaction term in the Hamiltonian is invariant
under that transformation, and the change of the Hamiltonian is
\begin{equation}
\begin{array}{l}
\delta H={\sum\limits_{ij}{Tr_{L\sigma }{\left[ {t_{ij}c_{i}^{+}\left( {%
U_{i}^{+}U_{j}-1}\right) c_{j}}\right] }}}=\delta _{1}H+\delta _{2}H \\
\delta _{1}H=\sin ^{2}{\frac{{\theta }}{{2}}}{\sum\limits_{k}{Tr_{L\sigma }}}%
{\left[ {\left( {t\left( {\bf k}{+}{\bf q}\right) -t\left( {\bf k}\right) }%
\right) c_{{\bf k}}^{+}c}_{{\bf k}}\right] } \\
\delta _{2}H={\frac{{1}}{{2}}}\sin \theta {\sum\limits_{ij}{Tr_{L}}}{\left[ {%
t_{ij}c_{i\downarrow }^{+}c_{j\uparrow }}\right] }\left( {\exp \left( {i{\bf %
qR}_{i}}\right) -\exp \left( {i{\bf qR}_{j}}\right) }\right)
\end{array}
\end{equation}

Consider further the case of small $\theta$, we can calculate the variation of the
total energy to lowest order in $\theta$ which corresponds to the first order
in $\delta _{1}H$ and the second order in $\delta _{2}H$:
\begin{eqnarray}
&&\delta E=\frac{\theta^{2}}{4}Tr_{L}\sum\limits_{\bf k}%
[t({\bf k}+{\bf q}) -t({\bf k})]\{n_{\bf k} +  \nonumber \\%
&&Tr_{\omega}[\gamma (k,q) G_{\downarrow }(k+q)[t({\bf k}+{\bf q})-t({\bf k})]%
G_{\uparrow }(k)]\},
\end{eqnarray}
where $n_{{\bf k}}=Tr_{L\sigma}{\left\langle {%
c_{{\bf k}}^{+}c_{{\bf k}}}\right\rangle}$,
$q,k$ are four-vectors with component $({\bf q},0)$ and $({\bf k},i\omega)$,
$\gamma$ is the three-leg vertex. Our main approximation is to neglect
the vertex corrections ($\gamma$ = 1). In this case the previous equation
takes the following form:
\begin{eqnarray}
&&\delta E= \frac{\theta ^{2}}{4}Tr_{L\omega}\sum\limits_{\bf k}%
[t({\bf k}+{\bf q})-t({\bf k})]* \\
&&G_{\downarrow}(k+q)[G_{\downarrow}^{-1}(k+q) - G_{\uparrow}^{-1}(k)+%
t({\bf k}+{\bf q}) -t({\bf k})] G_{\uparrow}(k) \nonumber
\label{deltaE}
\end{eqnarray}
Using the following consequence of the Dyson equation:
\begin{eqnarray}
t\left( {\bf k}{+}{\bf q}\right) -t\left( {\bf k}\right) =G_{\uparrow
}^{-1}\left( {k}\right) -G_{\downarrow }^{-1}\left( {k+q}\right) +\Sigma
_{\uparrow }\left( {E}\right) -\Sigma _{\downarrow }\left( {E}\right) \nonumber
\end{eqnarray}
one can rewrite the Eq.(\ref{deltaE}) in the form:
$\delta E={\frac{\theta^{2}}{4}}[J(0)-J({\bf q})]$
with the exchange integrals corresponding to Eq. (\ref{Jij}). We
conclude that the expression for $J_{ij}$ is accurate if the vertex
corrections can be neglected. Note that the limit of small {\bf q}
this neglecting can be justified rigorously, provided that the self-energy
and three-leg scalar vertex are local. This can be proven, e.g., using the
Ward-Takahashi identities \cite{HE}. Therefore, the expression for the stiffness
constant of the ferromagnet (Eq. (\ref{D21}))  appears to be exact in the framework
of DMFT \cite{berlin}.

%
%

\end{document}